\documentclass{aa}
\usepackage{graphicx}
\usepackage{txfonts}

\newcommand\mctwc[1]{\multicolumn{2}{c}{#1}}
\newcommand\mcfoc[1]{\multicolumn{4}{c}{#1}}

\begin{document}

\title{The dust-enshrouded microquasar candidate \object{AX~J1639.0$-$4642} = \object{IGR~J16393$-$4643}}

\author{J.~A. Combi\inst{1,2,3}
\and M. Rib\'o\inst{2}
\and I.~F. Mirabel\inst{2,4}
\and M. Sugizaki\inst{5}
}

\institute{
Instituto Argentino de Radioastronom\'{\i}a, C.C.5, (1894) Villa Elisa, Buenos Aires, Argentina\\
\email{pauletich@sinectis.com.ar}
\and Service d'Astrophysique, CEA Saclay, B\^at. 709, L'Orme des Merisiers, 91191 Gif-sur-Yvette, Cedex, France\\
\email{mribo@discovery.saclay.cea.fr; mirabel@discovery.saclay.cea.fr}
\and Departamento de F\'{\i}sica, Escuela Polit\'ecnica Superior, Univ. de Ja\'en, Virgen de la Cabeza 2, 23071 Ja\'en, Spain
\and Instituto de Astronom\'{\i}a y F\'{\i}sica del Espacio, CONICET,
C.C.67, Suc. 28, 1428 Buenos Aires, Argentina
\and Santa Cruz Institute for Particle Physics, University of California, 1156 High Street, Santa Cruz, CA 95064, USA\\
\email{sugizaki@scipp.ucsc.edu}
}


\offprints{J.~A. Combi,\\ \email{pauletich@sinectis.com.ar}}

\date{Received / Accepted}

\abstract{
We present a multiwavelength study of the field containing the unidentified
X-ray source \object{AX~J1639.0$-$4642}, discovered with the ASCA observatory
and recently detected with the IBIS telescope, onboard the INTEGRAL satellite, dubbed \object{IGR~J16393$-$4643}. The huge hydrogen column density
towards the source, the hard spectral index in the 0.7--10~keV band and its
flux variability suggest that the source is a High Mass X-ray Binary (HMXB)
enshrouded by dust. Our search reveals the presence of a non-thermal radio
counterpart within the X-ray error box. After a study of the broadband
emission from X-rays to the radio domain, we propose that
\object{AX~J1639.0$-$4642} is a dust-enshrouded Microquasar (MQ) candidate. In
addition, the X-ray source is well within the 95\% location contour of the
unidentified $\gamma$-ray source \object{3EG~J1639$-$4702}. The main
properties of \object{AX~J1639.0$-$4642}/\object{3EG~J1639$-$4702} are
consistent with those of two other MQs previously proposed to display
high-energy $\gamma$-ray emission.

}

\maketitle

\section{Introduction} \label{introduction}

Microquasars are X-ray binary systems with collimated relativistic jets
produced by the accretion of matter from a normal star to a compact object
such as a neutron star or a black hole (Mirabel \& Rodr\'{\i}guez
\cite{mirabel99}; Fender \cite{fender04}). This highly energetic process has
observable consequences from radio to hard X-rays and possibly up to
$\gamma$-ray energies (Paredes et~al. \cite{paredes00}). An interesting
property of microquasars is that, contrary to what happens in quasars, they
offer a unique opportunity to study accretion/ejection and related phenomena
on human timescales (Mirabel \& Rodr\'{\i}guez \cite{mirabel99}). Although the
total number of known MQ sources in the Galaxy is around 16 (Rib\'o
\cite{ribo03}), this number could increase up to $\sim$40 if, as it has been
suggested by Fender \& Hendry (\cite{fender00}), all Radio Emitting X-ray
Binaries (REXBs), are MQs. In any case, it is desirable to enlarge the MQ
population to allow meaningful statistical studies. To this end, Paredes
et~al. (\cite{paredes02}) have recently carried out a search for new REXBs at
galactic latitudes of $|b|<5\degr$, and although the first results were
promising (Rib\'o et~al. \cite{ribo02}), recent optical spectroscopic
observations (Mart\'{\i} et~al. \cite{marti04}) reveal that most of the
studied sources, if not all, are extragalactic quasars.

On the other hand, with the advent of the X-ray/$\gamma$-ray INTEGRAL
satellite the possibility to find new MQ candidates has increased. During the
last months, a few highly absorbed hard X-ray sources have been discovered
with the IBIS detector on the galactic plane towards the Norma spiral arm
(e.g. \object{IGR~J16318$-$4848}, Revnivtsev et~al. \cite{revnivetal03};
\object{IGR~J16320$-$4751}, Rodriguez et~al. \cite{rodriguez03}). These
sources present an unusually high intrinsic hydrogen column density up to $2
\times 10^{24}$~cm$^{-2}$, revealing a population of dust-enshrouded hard
X-ray sources, that are difficult to detect below 5~keV. Although at present
the nature of these sources remains unclear, it is believed that they could be
high mass X-ray binary systems (Revnivtsev \cite{revniv03}), and eventually
dust-enshrouded MQs.

With the aim to discover new MQs we have focused our attention on several yet
unexplored potential sources. In this paper, we present an in-depth study of
the unidentified X-ray source \object{AX~J1639.0$-$4642}, that was discovered
with ASCA (Sugizaki et~al. \cite{sugizaki01}), and show that it may be a MQ.
We present a re-analysis of the ASCA data in Sect.~\ref{asca}, we describe our
multiwavelength approach in Sect.~\ref{multi}, we discuss on the possible
origin of the detected broadband emission in Sect.~\ref{discussion} and we
state our conclusions in Sect.~\ref{conclusions}.

\section{Re-analysis of the ASCA data of \object{AX~J1639.0$-$4642}}
\label{asca}

The unidentified X-ray source \object{AX~J1639.0$-$4642} was discovered by the
Advanced Satellite for Cosmology and Astrophysics (ASCA) observatory during a
survey of the central region of the galactic plane, performed in the
0.7--10~keV energy range (Sugizaki et~al. \cite{sugizaki01}). The source is
superimposed to the galactic Norma spiral arm, and located at $(l, b) =
(337\fdg993, +0\fdg072)$, $(\alpha, \delta)_{\rm J2000.0} = (16^{\rm h}
39^{\rm m} 04\fs3, -46\degr 42\arcmin 47\arcsec)$ (90\% or 1.6$\sigma$
uncertainty of 1\arcmin). It presented a flux of $F_{\rm X (0.7-10~keV)}$=
$(19.2 \pm 2.2) \times 10^{-12}$ erg~cm$^{-2}$~s$^{-1}$ (and variable X-ray
emission with a confidence $\geq$ 99\%). Its spectrum was fit with a power-law
with a very hard photon index $\Gamma=-0.01^{+0.66}_{-0.60}$ and a poorly
constrained hydrogen column density of $N_{\rm
H}=12.82^{+8.58}_{-6.88}\times10^{22}~$cm$^{-2}$. The source has recently been
detected by the IBIS telescope onboard the INTEGRAL satellite (Malizia et~al.
\cite{malizia04}).


\subsection{Observations and data reduction}

In order to investigate in more detail the properties of this X-ray source, we
have re-analyzed the original ASCA data. The observations were performed on
1997 September 10, with two different pointings in the overlapped region of
successive GIS fields of view. The first pointed observation (OBS\#1) started
on 07:01:34 (UT) and had an exposure time of 11~ks, while the second one
(OBS\#2) started on 12:41:50 (UT) and lasted 6~ks. The source was located at
an off-axis angle of $20$\arcmin\ and $14$\arcmin\ from the optical axis of
the X-ray telescope in OBS\#1 and OBS\#2, respectively.

The data reduction and the extraction of events was performed based on the
standard procedure, as described in Sugizaki et~al. (\cite{sugizaki01}). The
source events were extracted from a circular region of 3~mm radius on the GIS
detector around the source position. Since the background is angle-dependent,
it was collected from a source-free region with the same off-axis angle as the
source. The data of the GIS-2 and the GIS-3 were summed to gain photon
statistics.

\subsection{Lightcurve}

\begin{figure}
\resizebox{\hsize}{!}{\includegraphics[angle=270]{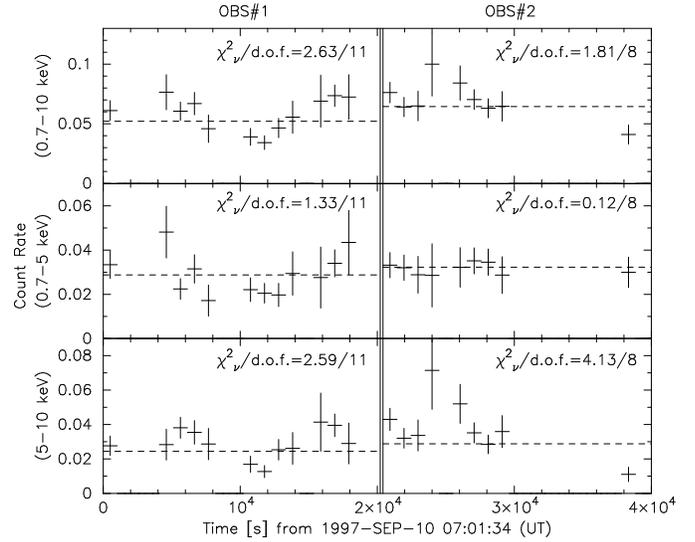}}
\caption{Background subtracted GIS lightcurves of \object{AX~J1639.0$-$4642} in 0.7--10~keV (top), 0.7--5~keV (middle) and 5--10~keV (bottom) throughout OBS\#1 (0--2$\times 10^{4}$~s) and OBS\#2 (2--4$\times 10^{4}$~s). Error bars represent 1$\sigma$ statistical uncertainty. Dashed lines represent the best-fit constant models. Reduced chi-squared ($\chi^2_\nu$) and degree of freedom (d.o.f.) of the best-fit models are shown together. The different average count rates between OBS\#1 and OBS\#2 are explained by the change of the source position on the detector.}
\label{fig:lc}
\end{figure}

In order to examine the time variability of \object{AX~J1639.0$-$4642}, we
extracted lightcurves and fitted them with a constant model. We show in
Fig.~\ref{fig:lc} the lightcurves obtained throughout the two pointed
observations in the total energy band 0.7--10~keV, and in the energy bands of
0.7--5 and 5--10~keV. The best-fit constant model and the reduced chi-squared
($\chi^2_\nu$) of each lightcurve are also shown in Fig.~\ref{fig:lc}.

In the 0.7--5~keV range, the values of $\chi^2_\nu$ are within the 90\%
confidence limits, thus the flux can be considered to be constant. However, in
the 5--10~keV range, the $\chi^2_\nu$ values are over the 99\% confidence
limits, which implies that the flux is significantly variable. Nevertheless,
the photon statistics are not good enough to conclude that the source has
larger variability in the hard band than in the soft band.

\subsection{Spectral analysis}

\begin{figure}
\resizebox{\hsize}{!}{\includegraphics[angle=270]{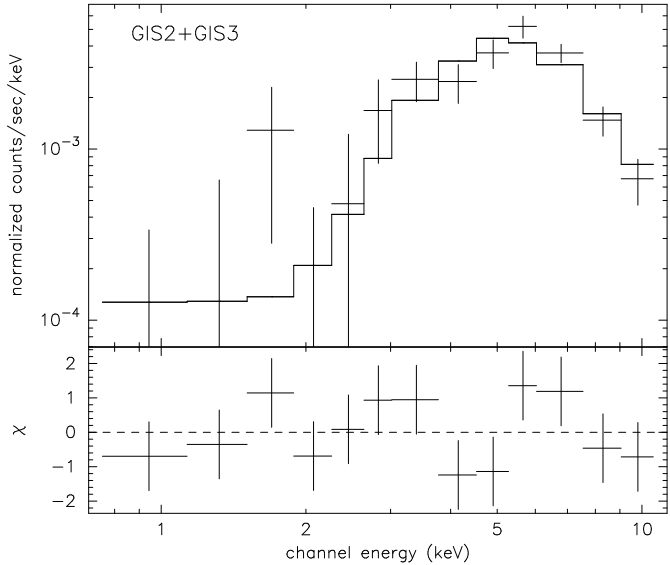}}
\caption{GIS spectrum of \object{AX~J1639.0$-$4642} with the best-fit power-law model with an absorption of neutral matter with solar abundance (top), and the residuals (bottom). Error bars represent 1$\sigma$ statistical uncertainty.}
\label{fig:spec}
\end{figure}

The spectral parameters of \object{AX~J1639.0$-$4642} in the ASCA galactic
faint source survey in Sugizaki et~al. (\cite{sugizaki01}) were based on the
OBS\#1 data. However, in OBS\#1 the source was observed at a large off-axis
angle of 20\arcmin, where the response function of the X-ray telescope is not
well calibrated. Therefore, we have re-analyzed the X-ray spectrum using the
OBS\#2 data, when the source was observed with a lower off-axis angle of
14\arcmin. All spectral uncertainties represent 90\% confidence limits,
hereafter. We show in Fig.~\ref{fig:spec} the obtained X-ray spectrum in
OBS\#2. We first fitted the spectrum with a power-law model with an absorption
of neutral matter of solar abundance. The obtained best-fit parameters were
photon index, $\Gamma=-0.28_{-0.56}^{+1.1}$, absorption hydrogen column
density, $N_{\rm H}=11_{-7}^{+17} \times 10^{22}$ cm$^{-2}$, and flux in the
0.7--10~keV band, $F_{\rm 0.7-10\, keV}=(11 \pm 1.2) \times 10^{-12}$
erg~cm$^{-2}$~s$^{-1}$. These values are consistent with those from OBS\#1 in
the ASCA galactic faint X-ray catalog, although the flux is somehow smaller.
The obtained value for $\Gamma$$\sim$0 is very poorly constrained, and is
certainly smaller than the typical values of $\Gamma$$>$1 found in other X-ray
binaries. Therefore, in order to estimate a more realistic absorption column
density, we next fitted the spectrum with the power-law model fixing the
photon index $\Gamma$ to be 1, which is typical of hard X-ray sources such as
HMXBs. The obtained acceptable range of $N_{\rm H}$ in 90\% confidence was
$30_{-9}^{+12}\times 10^{22}$ cm$^{-2}$. We also attempted to make a fit with
a blackbody as an example of thermal spectral models, and derived a
temperature of $kT=5.5_{-2.7}^{+40}$~keV at the best-fit. However, the thermal
spectral model is unlikely because the source has been detected by INTEGRAL
ISGRI in the 15--40~keV band. Unfortunately, the photon statistics of the ASCA
observations are not good enough to exclude any model with a confidence
greater than 99\%. The results of the spectral fitting are summarized in
Table~\ref{table:spec_para}.

\begin{table}
\begin{center}
\caption[]{Summary of spectral fitting results. All errors represent the 90\% confidence limits of statistical uncertainty.}
\label{table:spec_para}
\begin{tabular}{lccc}
\hline \hline \noalign{\smallskip}
                              & \multicolumn{3}{c}{Model}\\
Parameter                     & PL       & PL ($\Gamma=1$:fixed) & BB\\
\noalign{\smallskip} \hline \noalign{\smallskip}
$\Gamma$ or $kT$$^{a}$        & $-0.28_{-0.56}^{+1.1}$ & 1:fix           & $5.5_{-2.7}^{+40}$ \\
$N_{\rm H}$$^{b}$             & $ 11_{-7}^{+17}$       & $30_{-9}^{+12}$ & $10_{-6}^{+13}$ \\
$F_{\rm 0.7-10\, keV}$$^{c}$ & 1.1$\pm$0.13            & 0.99$\pm$0.12   & 1.1$\pm$0.13 \\
\noalign{\smallskip} \hline \noalign{\smallskip}
$\chi^2_\nu$/d.o.f.          & 1.10/10                 & 1.33/11         & 1.04/10 \\
\noalign{\smallskip} \hline
\end{tabular}
\end{center}
$^{a}$ $kT$ in units of keV.
$^{b}$ Absorption hydrogen column density in units of $10^{22}$~cm$^{-2}$.
$^{c}$ Flux in units of 10$^{-11}$ erg~cm$^{-2}$~s$^{-1}$.
\end{table}

\section{Multiwavelength study of the \object{AX~J1639.0$-$4642} field} \label{multi}

\begin{figure}[t!]
\resizebox{\hsize}{!}{\includegraphics{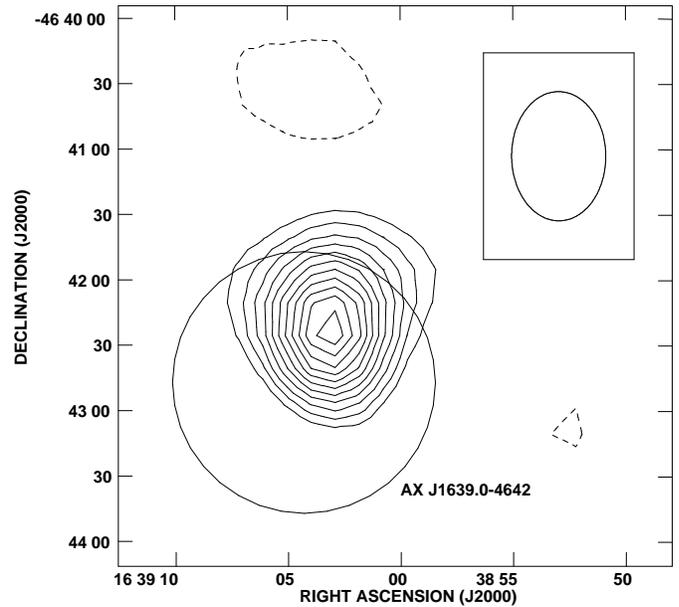}}
\caption[]{Contour image of the MGPS data obtained with MOST at 843~MHz on 1992 April 16. The image size is 4\arcmin$\times$4\arcmin. The radio source MOST~J1639.0$-$4642 is well within the 90\% uncertainty error circle of the X-ray source \object{AX~J1639.0$-$4642}. Contours are $-$2, 2, 3, 4, 5, 6, 7, 8, 9, 10, 11 and 12 times the rms noise level of 10~mJy. The ellipse on the top right corner is the convolving beam of 59.2$\times$43.0 arcsec in PA=0\degr.}
\label{fig:most}
\end{figure}

With the aim of finding any possible radio counterpart within the location
error box of the X-ray source, we have used the Molonglo Galactic Plane Survey
(MGPS) conducted at 843~MHz with the Molonglo Observatory Synthesis Telescope
(MOST) (Green et~al. \cite{green99}). Observations of the region of
interest were conducted on 1990 June~9 and on 1992 April 16. We show in
Fig.~\ref{fig:most} the MOST contour image of the surroundings of
\object{AX~J1639.0$-$4642} obtained during the second run. A bright radio
source (hereafter MOST~J1639.0$-$4642) can be clearly seen well within the
error box of the X-ray source. Although the source appears somehow extended,
it is point-like when considering the beam size and the presence of nearby
sources. We have fitted the images of the two epochs with Gaussian functions
and we have obtained peak flux densities of $140\pm14$ and
$132\pm11$~mJy~beam$^{-1}$, respectively. The position of the radio source in
galactic and equatorial coordinates is $(l, b) = (337\fdg995, +0\fdg079)$,
$(\alpha, \delta)_{\rm J2000.0} = (16^{\rm h} 39^{\rm m} 03\fs0, -46\degr
42\arcmin 20\arcsec)$ (3$\sigma$ uncertainty of 15\arcsec). We note that
although the formal errors in right ascension and declination of MGPS are
$1\arcsec\times1\arcsec\csc|\delta|$ for sources stronger than 20~mJy,
sidelobes from nearby strong sources can make these uncertainties much higher.
In fact, the peak and the fitted Gaussian center positions in the two epochs
change by several arcseconds. Therefore, we have estimated the 1$\sigma$ error
radius in position to be 5\arcsec.





In order to search for the radio source at higher frequencies, we have
inspected the data from the 4.85~GHz PMN single dish survey (Condon et~al.
\cite{condon93}). We applied to these data an additional Gaussian filtering
process (see Combi et~al. \cite{combi98} for details) to remove the background
diffuse radiation on scales larger than 8\arcmin. After several iterations no
radio source was found up to a 3$\sigma$ level of $\sim$15~mJy.

\begin{figure}[t!]
\resizebox{\hsize}{!}{\includegraphics{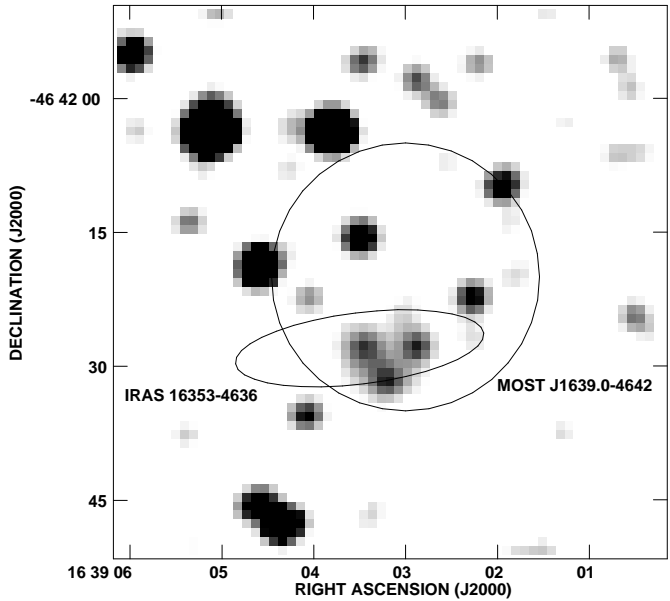}}
\caption[]{2MASS $K_s$-band 60\arcsec$\times$60\arcsec\ image of the environment of the radio source MOST~J1639.0$-$4642, whose 3$\sigma$ position error circle is shown, together with the 2$\sigma$ position error ellipse of the far infrared source \object{IRAS~16353$-$4636}. Several NIR sources from the 2MASS catalog are possible counterparts of both sources.}
\label{fig:2mass}
\end{figure}

At near infrared (NIR) wavelengths we have inspected the 2~Micron All Sky
Survey (2MASS, Cutri et~al. \cite{cutri03}), and found 10 sources in the
3$\sigma$ error circle in position of MOST~J1639.0$-$4642, some of them
visible in the $K_s$-band image shown in Fig.~\ref{fig:2mass}. At optical
wavelengths we have queried the USNO-B1.0 catalog (Monet et~al.
\cite{monet03}), and found 6 sources in the same error circle, being 3 of them
located within 1 arcsecond of 2MASS sources.




At the far infrared part of the spectrum, from 12 to 100 microns, we have
found that the source \object{IRAS~16353$-$4636}\footnote{{\tt
http://cdsweb.u-strasbg.fr/viz-bin/Cat?II/125}} lies inside the error box of
the X-ray source. This infrared source is located at $(l, b) = (337\fdg995,
+0\fdg077)$, $(\alpha, \delta)_{\rm J2000.0} = (16^{\rm h} 39^{\rm m} 03\fs5,
-46\degr 42\arcmin 28\arcsec)$ (95\% or 2$\sigma$ uncertainty ellipse of
14\arcsec$\times$4\arcsec\ in PA=97\degr). The infrared fluxes at 12, 25, 60
and 100 microns are 12.5$\pm$0.6, 73.3$\pm$3.7, $<$806 (3$\sigma$ upper limit)
and 2230$\pm$330~Jy, respectively. After correction due to the slope of the
spectrum, these fluxes become 13.5$\pm$0.7, 80.5$\pm$4.0, $<$806 (3$\sigma$
upper limit) and 2210$\pm$330~Jy, respectively. This source overlaps the
southern part of the MOST~J1639.0$-$4642 3$\sigma$ position error circle, and
its uncertainty ellipse in position contains several 2MASS sources, as can be
seen in Fig.~\ref{fig:2mass}.

The X-ray source \object{AX~J1639.0$-$4642} has been recently re-discovered at
higher energies with the IBIS telescope onboard the INTEGRAL satellite, dubbed
\object{IGR~J16393$-$4643} (Malizia et~al. \cite{malizia04}). It is located at
$(l, b) = (338\fdg02, +0\fdg04)$, $(\alpha, \delta)_{\rm J2000.0} = (16^{\rm
h} 39^{\rm m} 18^{\rm s}, -46\degr 43\arcmin 00\arcsec)$ (90\% or 1.6$\sigma$
uncertainty of 2\arcmin). We note that although this error circle does not
include neither the center of \object{AX~J1639.0$-$4642} nor the source
MOST~J1639.0$-$4642, both sources fall within the 2$\sigma$ error circle of
\object{IGR~J16393$-$4643}. This source shows an average flux of $F_{\rm X
(20-100~keV)}\simeq5\times 10^{-11}$ erg~cm$^{-2}$~s$^{-1}$, and presents a
factor of 2--3 flux variability on timescales of months.

\begin{figure}[t!]
\resizebox{\hsize}{!}{\includegraphics{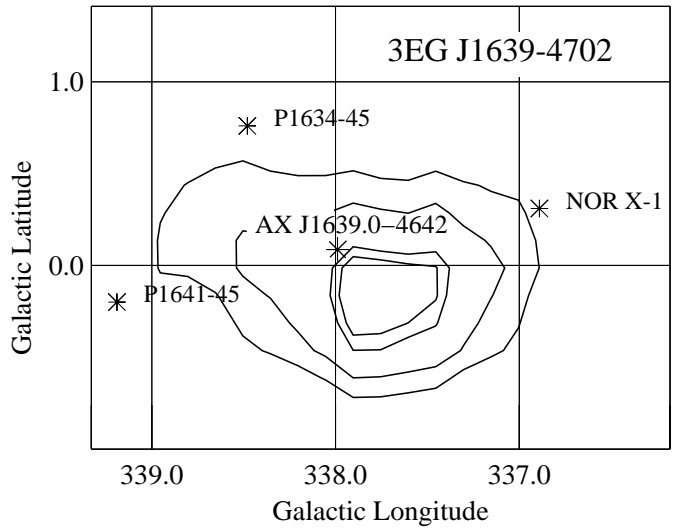}}
\caption[]{Gamma-ray probability contours (50\%, 68\%, 95\%, and 99\%, from inside to outside) of the unidentified source \object{3EG~J1639$-$4702}. The position of the X-ray source \object{AX~J1639.0$-$4642}, well within the 95\% probability contour, is marked with an asterisk.}
\label{fig:egret}
\end{figure}

Finally, it is interesting to point out that \object{AX~J1639.0$-$4642} lies
inside the 95\% location contour of the unidentified $\gamma$-ray source
\object{3EG~J1639$-$4702} (Hartman et~al. \cite{hartman99}), as can be seen in
Fig.~\ref{fig:egret}. The source is located at $(l, b) = (337\fdg75,
-0\fdg15)$, $(\alpha, \delta)_{\rm J2000.0} = (16^{\rm h} 39^{\rm m} 07^{\rm
s}, -47\degr 02\arcmin 24\arcsec)$, and has a 2$\sigma$ radius of 0\fdg56. Its
$\gamma$-ray flux is $(53.2\pm8.7)\times10^{-8}$~photon~cm$^{-2}$~s$^{-1}$,
and presents a steep $\gamma$-ray spectral index of $\Gamma=2.5\pm0.18$.


The positions of all these sources are summarized in Table~\ref{table:coord}.

\begin{table*}[t!]
\begin{center}
\caption[]{Galactic and equatorial coordinates and position uncertainties of the sources discussed in the text.}
\label{table:coord}
\begin{tabular}{lr@{\fdg}l@{~~~}r@{\fdg}l@{~~~~}r@{$^{\rm h}$\,}r@{$^{\rm m}$\,}r@{\fs}l@{~~~}r@{\degr\,}r@{\arcmin\,}r@{\arcsec}ll}
\hline \hline \noalign{\smallskip}
Source name                & \mctwc{$l$}  & \mctwc{$b$}  & \mcfoc{$\alpha_{\rm (J2000.0)}$} & \mcfoc{$\delta_{\rm (J2000.0)}$} & Pos. uncertainty \\ 
\noalign{\smallskip} \hline \noalign{\smallskip}
\object{AX~J1639.0$-$4642} & 337&993   & +0&072   & 16&39&04&3    & $-$46&42&47&   & 1\arcmin\ (90\% or 1.6$\sigma$)\\ 
MOST~J1639.0$-$4642        & 337&995   & +0&079   & 16&39&03&0    & $-$46&42&20&   & 15\arcsec\ (99.7\% or 3$\sigma$) \\ 
\object{IRAS~16353$-$4636} & 337&995   & +0&077   & 16&39&03&5    & $-$46&42&28&   & 14\arcsec$\times$4\arcsec\ in PA=97\degr\ (95\% or 2$\sigma$)\\
\object{IGR~J16393$-$4643} & 338&02    & +0&04    & 16&39&18&     & $-$46&43&00&   & 2\arcmin\ (90\% or 1.6$\sigma$) \\ 
\object{3EG~J1639$-$4702}  & 337&75    & $-$0&15  & 16&39&07&     & $-$47&02&24&   & 0\fdg56 (95\% or 2$\sigma$)\\ 
\noalign{\smallskip} \hline
\end{tabular}
\end{center}
\end{table*}

\section{Discussion} \label{discussion}

\subsection{On the nature of \object{AX~J1639.0$-$4642}}


At radio wavelengths, the detection at 843~MHz (observations on 1990 June~9
and 1992 April 16) and the non-simultaneous non-detection at 4.85~GHz
(observations in 1990 January-March), imply that either the source is highly
variable on timescales of months or that it has a steep spectral index
$\alpha<-1$ (defined in such a way that S$_{\nu} \propto \nu^{+\alpha}$). In
both cases, a non-thermal origin of the radio emission is clearly favored.
Therefore, the radio source detected inside the error box of
\object{AX~J1639.0$-$4642} could be the manifestation of synchrotron emission
from a jet.

The flux variability of \object{AX~J1639.0$-$4642} shown in Fig.~\ref{fig:lc}
and the X-ray spectrum shown in Fig.~\ref{fig:spec} suggest that it is a HMXB,
as already pointed out by Sugizaki et~al. (\cite{sugizaki01}). In addition,
\object{AX~J1639.0$-$4642} appears to be a persistent X-ray source, since it
was detected by ASCA and INTEGRAL, which supports again a HMXB. Finally, the
factor of 2--3 flux variability on timescales of months of
\object{IGR~J16393$-$4643} also points towards a HMXB.

The obtained value of $N_{\rm H}$ above 10$^{23}$~cm$^{-2}$ is reminiscent of
the population of persistent and highly absorbed IGR sources detected by
INTEGRAL towards the Norma spiral arm, a region of high concentration of
atomic and molecular gas, and OB stars. All these IGR sources (e.g.
\object{IGR~J16318$-$4848}, Revnivtsev et~al. \cite{revnivetal03};
\object{IGR~J16320$-$4751}, Rodriguez et~al. \cite{rodriguez03}) present
rather similar hard X-ray spectra, which is typical of HMXB systems. The most
interesting characteristic of these sources is the variable hydrogen column
density measured in different observations, a fact that supports an intrinsic
absorption of the sources.

In this context, we can roughly estimate the visual extinction $A_V$ along the
line of sight of \object{AX~J1639.0$-$4642} through its relation to $N_{\rm
H}$ obtained from the X-ray spectrum. Using the formula by Predehl \& Schmitt
(\cite{predehl95}), $[A_V/{\rm mag}]=0.56\,[N_{\rm H}/10^{21}{\rm
cm}^{-2}]+0.23$, and the $N_{\rm H}$ values up to 90\% confidence limits from
the PL model in Table~\ref{table:spec_para}, we obtain extinction values in
the range $\sim$20--160 magnitudes. Using the PL model with $\Gamma$=1, $A_V$
ranges between $\sim$120--240 magnitudes.

The higher estimated values of $A_V$ could be consistent with the binary
system lying inside a molecular cloud, a possibility supported by the presence
of the CO molecular cloud located between 8--12~kpc (Fig.~17 in Torres et~al.
\cite{torres03}). Alternatively, the binary system could be enshrouded by a
dense dust envelope, as it has been proposed for the IGR sources discussed
above. In any of these cases, we would not expect to detect the NIR/optical
counterpart.

However, variable absorption has been measured in \object{IGR~J16318$-$4848}
through RXTE and XMM observations on timescales of months (Revnivtsev
\cite{revniv03}). Moreover, in a study of the properties of the absorbing and
line emitting material in this source, Matt \& Guainazzi (\cite{matt03}) have
observed that the iron $K_{\alpha}$ line varies on timescales as short as
1000~s, implying that the emitting region should have the size of the binary
system. This is consistent with a picture in which the absorbing material is
due to the stream flowing through the Lagrangian point to form an accretion
disk. Although the ASCA data is consistent with a constant X-ray flux below
5~keV, the poor statistics below 2--3~keV do not prevent a similar situation
to occur in \object{AX~J1639.0$-$4642}. In such a case, the absorbing material
would be placed within the binary system and very close to the compact object,
and we could expect a low value of the extinction towards the companion star,
which could be eventually detected.

Although we do not have spectrophotometric information of a NIR/optical
counterpart in order to derive a distance to the source, assuming that it is
located in the Scutum-Crux or in the Norma spiral arms, the galactic longitude
of $l=338\degr$ leads to distances in the range between 3 and 13~kpc.


All these results show that the source \object{AX~J1639.0$-$4642}
(=\object{IGR~J16393$-$4643}) clearly deserves further attention. High
resolution multifrequency radio observations have recently been conducted to
assess the expected non-thermal nature of the emission, to infer physical
parameters, to look for eventually extended emission and to provide a better
position estimate of the source, to check if it is positionally coincident
with \object{IRAS~16353$-$4636} and to look for a NIR counterpart. Further NIR
spectroscopy of an eventual counterpart would allow to obtain the spectral
type of the optical companion, which would allow to estimate the distance to
the source and the luminosity in the different spectral domains. X-ray and
soft $\gamma$-ray observations with currently orbiting satellites will allow
to obtain a much better position estimate of \object{AX~J1639.0$-$4642} and
will constrain the spectral parameters of the source, such as the photon index
and specially the hydrogen column density and look for its eventual
variability.

\subsection{Microquasars as possible counterparts of unidentified EGRET sources}

As reported in the previous section, \object{AX~J1639.0$-$4642} could be
associated with the unidentified high-energy $\gamma$-ray source
\object{3EG~J1639$-$4702}. Although Torres et~al. (\cite{torres01b}) have
found three radio pulsars inside the 95\% confidence contour of the
$\gamma$-ray source, its possible variability (index $I=1.95$) and steep
photon index ($\Gamma_\gamma=2.50\pm0.18$) do not seem to agree with a pulsar
origin. Similarly, these properties would rule out an association with the
three SNRs found in the 95\% confidence contour (Torres et~al.
\cite{torres03}). Moreover, no identified blazar has been found within the
$\gamma$-ray contours. Therefore, we suggest that the microquasar candidate
\object{AX~J1639.0$-$4642}/MOST~J1639.0$-$4642 (=\object{IGR~J16393$-$4643})
is the counterpart of \object{3EG~J1639$-$4702}.

\begin{table*}[t!]
\begin{center}
\caption[]{Properties of the three $\gamma$-ray sources and the proposed X-ray/optical/radio counterparts. In the cases of the microquasars \object{LS~5039} and \object{LS~I~+61~303} the luminosity intervals correspond to intrinsic variability of the sources at the corresponding distances, while in the case of \object{AX~J1639.0$-$4642} reflect the range of assumed possible distances.}
\label{table:egret}
\begin{tabular}{@{}c@{~~~}c@{~~~}c@{~~~}c@{~~~}c@{~~~}c@{~~~}c@{~~~}c@{~~~}c@{~~~}c@{}}
\hline \hline \noalign{\smallskip}
$\gamma$-ray source& $\Gamma_{\gamma}$$^{a}$ & $I$\,$^{b}$ & $L_{\gamma (>100~{\rm MeV})}$$^{c}$  & X-ray source & $L_{\rm X (0.7-10~keV)}$       & $L_{\rm radio (0.1-100~GHz)}$            & Spectral            & $P_{\rm orb}$  & $d$ \\
                   &                         &           & (erg~s$^{-1}$)                         &              & (erg~s$^{-1}$)                 & (erg~s$^{-1}$)                           & type                & (days)         & (kpc)\\
\noalign{\smallskip} \hline \noalign{\smallskip}
\object{3EG~J1824$-$1514} & 2.19$\pm$0.18 & 3.00 & ~~~~~~3.6$\times$10$^{35}$  & \object{LS~5039}           & ~(0.5--5)$\times$10$^{34}$\,$^{d}$  & ~~~~~$\sim$1.0$\times$10$^{31}$\,$^{e}$  & ON6.5V((f))\,$^{f}$ & ~~4.4\,$^{f}$  & \,2.9\,$^{g}$ \\
\object{3EG~J0241$+$6103} & 2.21$\pm$0.07 & 1.31 & ~~~~~\,3.1$\times$10$^{35}$ & \object{LS~I~+61~303}      & ~~~~(1--6)$\times$10$^{34}$\,$^{h}$ & ~\,(1--17)$\times$10$^{31}$\,$^{i}$      & B0Ve\,$^{j}$        &  26.5\,$^{k}$  & 2.0\,$^{l}$ \\
\object{3EG~J1639$-$4702} & 2.50$\pm$0.18 & 1.95 & (3--50)$\times$10$^{35}$    & \object{AX~J1639.0$-$4642} & (2--40)$\times$10$^{34}$            & (0.8--16)$\times$10$^{31}$\,$^{m}$       & ?                   & ?              & 3--13 ?\\
\noalign{\smallskip} \hline
\end{tabular}
\end{center}
$^{a}$ Hartman et~al. (\cite{hartman99}); $^{b}$ Torres et~al. (\cite{torres01a}); $^{c}$ computed using the photon fluxes and indexes from Hartman et~al. (\cite{hartman99}); $^d$ Reig et~al. (\cite{reig03}); $^e$ Mart\'{\i} et~al. (\cite{marti98}), Rib\'o et~al. (\cite{ribo99}); $^f$ McSwain et~al. (\cite{mcswain04}); $^g$ Rib\'o et~al. (\cite{ribo02}); $^h$ Paredes et~al. (\cite{paredes97}); $^i$ computed from the values quoted in Strickman et~al. (\cite{strickman98}) ; $^j$ Hutchings \& Crampton (\cite{hutchings81}) ; $^k$ Gregory (\cite{gregory02}); $^{l}$ Frail \& Hjellming (\cite{frail91}); $^m$ computed assuming $\alpha=-1$.
\end{table*}

The possibility of MQ being $\gamma$-ray emitters has been suggested by
Paredes et~al. (\cite{paredes00}), who proposed the association between the
HMXB \object{LS~5039} and the unidentified EGRET source
\object{3EG~J1824$-$1514}. In their scenario the $\gamma$-rays are produced by
Inverse Compton upscattering of stellar UV photons by the non-thermal
relativistic electron population that later on will produce the detected radio
emission. On the other hand, the X-ray binary system \object{LS~I~+61~303} has
been associated with the EGRET source \object{3EG~J0241$+$6103} (Tavani et~al.
\cite{tavani96}; Kniffen et~al. \cite{kniffen97}; Strickman et~al.
\cite{strickman98}; Harrison et~al. \cite{harrison00}), and recently Massi
et~al. (\cite{massi01}, \cite{massi04}) have revealed its MQ nature. If the MQ
nature of \object{AX~J1639.0$-$4642} is confirmed, it could be the third MQ
source related to a high-energy gamma-ray source detected by the EGRET
telescope. We quote the basic properties of these three $\gamma$-ray sources
in Table~\ref{table:egret}, together with the properties of the proposed X-ray
counterparts. It should be noted that \object{LS~5039} and
\object{LS~I~+61~303}, and probably \object{AX~J1639.0$-$4642}, have massive
optical companions, which provide a stellar UV photon field. On the other
hand, the compact object appears to be compatible with a neutron star in the
cases of \object{LS~5039} (McSwain et~al. \cite{mcswain04}) and
\object{LS~I~+61~303} (Hutchings \& Crampton \cite{hutchings81}). In addition,
it is interesting to point out that the luminosities obtained in each spectral
domain are very similar in all three sources, specially for the shorter
distances to \object{AX~J1639.0$-$4642}, giving support to the idea that all
of them have similar emission processes.

In this context, Kaufman Bernad\'o et~al. (\cite{kaufman02}) have recently
presented a model for galactic $\gamma$-ray emission based on
precessing MQ with a jet pointing close to the line of sight (or
microblazars).
Depending on the parameters introduced in the model, such MQs can explain
sources with $\gamma$-ray luminosities in the range of 10$^{34}$--10$^{38}$
erg~s$^{-1}$ in the observer's frame, perfectly consistent with the values
quoted in Table~\ref{table:egret} for all three sources. More recently, Romero
et~al. (\cite{romero03}) have introduced a new mechanism for the generation of
high-energy $\gamma$-rays in MQs, based on hadronic interactions occurring
outside the coronal X-ray region. The three gamma-ray sources under
consideration could be part of this hadronic microblazar population.

Observations with the future missions AGILE and GLAST will confirm or reject
the proposed association between \object{AX~J1639.0$-$4642} and the
high-energy $\gamma$-ray source \object{3EG~J1639$-$4702}.

\section{Conclusions} \label{conclusions}

In this paper we have re-analyzed the ASCA data of the X-ray source
\object{AX~J1639.0$-$4642} (=\object{IGR~J16393$-$4643}) and presented a
multiwavelength study of emission in the direction of the source from radio to
$\gamma$-rays. The huge hydrogen column density, hard spectral index and flux
variability suggest that the source is a dust-enshrouded HMXB system. We have
found a non-thermal radio source inside its error circle that could be the 
manifestation of synchrotron emission from a jet, hence supporting its
microquasar nature. On the other hand, apart from \object{AX~J1639.0$-$4642},
there are no other known candidate sources within the 95\% location contours
of \object{3EG~J1639$-$4702} capable to generate the GeV emission detected by
EGRET. All these facts suggest that \object{AX~J1639.0$-$4642}
(=\object{IGR~J16393$-$4643}) is probably a dust-enshrouded galactic
microquasar with high-energy $\gamma$-ray emission.


\begin{acknowledgements}


We acknowledge Sylvain Chaty and Gustavo E. Romero for useful discussions, and
Paula Benaglia for useful information prior to publication. J.A.C. was
supported by CONICET (under grant PIP N$^{\circ}$ 0430/98). J.A.C. is a
researcher of the programme {\em Ram\'on y Cajal} funded by the University of
Ja\'en and Spanish Ministery of Science and Technology. J.A.C. is very
grateful to staff of the Service d'Astrophysique (CEA Saclay) and of
Universidad de Ja\'en, where his research for this project was carried out.
M.R. acknowledges support by a Marie Curie Fellowship of the European
Community programme Improving Human Potential under contract number
HPMF-CT-2002-02053. M.R. also acknowledges partial support by DGI of the
Ministerio de Ciencia y Tecnolog\'{\i}a (Spain) under grant AYA2001-3092, as
well as partial support by the European Regional Development Fund
(ERDF/FEDER).
This research has made use of the NASA's Astrophysics Data System Abstract
Service, of the SIMBAD database, operated at CDS, Strasbourg, France, and of
the NASA/IPAC Extragalactic Database (NED) which is operated by the Jet
Propulsion Laboratory, California Institute of Technology, under contract with
the National Aeronautics and Space Administration. The Digitized Sky Survey
was produced at the Space Telescope Science Institute under U.S. Government
grant NAG~W-2166.
This publication makes use of data products from the Two Micron All Sky
Survey, which is a joint project of the University of Massachusetts and the
Infrared Processing and Analysis Center/California Institute of Technology,
funded by the National Aeronautics and Space Administration and the National
Science Foundation.

\end{acknowledgements}

\end{document}